\documentclass[%
 reprint,
 superscriptaddress,
 amsmath,amssymb,
 aps,
]{revtex4-1}

\usepackage{graphicx}
\usepackage{dcolumn}
\usepackage{bm}
\usepackage{xcolor}
\usepackage{url}
\usepackage{longtable}

\usepackage{subfigure}
\usepackage{multirow}

\begin{document}


\title{Extraction of $\omega$n, $\omega$p and $\phi$N scattering lengths from $\omega$ and $\phi$ differential photoproduction cross sections on the deuterium target}

\author{Chengdong Han}
\email{chdhan@impcas.ac.cn}
\affiliation{Institute of Modern Physics, Chinese Academy of Sciences, Lanzhou 730000, China}
\affiliation{University of Chinese Academy of Sciences, Beijing 100049, China}

\author{Wei Kou}
\email{kouwei@impcas.ac.cn}
\affiliation{Institute of Modern Physics, Chinese Academy of Sciences, Lanzhou 730000, China}
\affiliation{University of Chinese Academy of Sciences, Beijing 100049, China}

\author{Rong Wang}
\email{rwang@impcas.ac.cn}
\affiliation{Institute of Modern Physics, Chinese Academy of Sciences, Lanzhou 730000, China}
\affiliation{University of Chinese Academy of Sciences, Beijing 100049, China}

\author{Xurong Chen}
\email{xchen@impcas.ac.cn (Corresponding author)}
\affiliation{Institute of Modern Physics, Chinese Academy of Sciences, Lanzhou 730000, China}
\affiliation{University of Chinese Academy of Sciences, Beijing 100049, China}
\affiliation{Guangdong Provincial Key Laboratory of Nuclear Science, Institute of Quantum Matter, South China Normal University, Guangzhou 510006, China}


\date{\today}

\begin{abstract}
  In this study, we try to extract $\omega n$, $\omega p$ and $\phi$N scattering lengths from the differential cross-section data of near-threshold $\omega$
  and $\phi$ photoproductions not only to the energy at threshold t$_{thr}$, but also in t to t = 0.
  The incoherent data of $\omega$ and $\phi$ photoproductions on deuterium target for scattering length extractions
  are provided by CBELSA/TAPS Collaboration and LEPS Collaboration respectively,
  where the deuteron is usually approximated as a weakly bound state of a quasi-free neutron plus a quasi-free proton.
  Under the Vector Meson Doninance model and the assumption of energy independence of the differential cross-section, 
  we obtained the $|\alpha_{\omega n^{*}}|$ and $|\alpha_{\omega p^{*}}|$ of loosely bound neutron and proton in the deuteron to be
  0.709 $\pm$ 0.078 fm and 0.621 $\pm$ 0.022 fm respectively at threshold t$_{thr}$ (1.258 $\pm$ 0.130 fm
  and 1.056 $\pm$ 0.032 fm respectively in t to t = 0) for the first time by fitting the near-threshold data of
  $\gamma$d $\rightarrow$ $\omega$n(p) and $\gamma$d $\rightarrow$ $\omega$p(n).
  For a comparison study, we also extracted the scattering length $|\alpha_{\omega p}|$ of the free proton from
  the $\omega$ photoproduction on the hydrogen target by CBELSA/TAPS Collaboration.
  With the near-threshold and incoherent $\phi$ photoproduction data of $\gamma$d $\rightarrow$ $\phi$pn by LEPS Collaboration, we obtained
  scattering length $|\alpha_{\phi N}|$ of the bound nucleon inside the deuteron to be 0.109 $\pm$ 0.008 fm
  at threshold t$_{thr}$ and 0.276 $\pm$ 0.010 fm in t to t = 0, respectively.
  The obtained $|\alpha_{\omega p}|$ and $|\alpha_{\phi N}|$ are basically in agreement with the experimental indications and the theoretical predictions
  within the current uncertainties.
\end{abstract}

\maketitle


\section{Introduction}
\label{introduction}
Vector meson-nucleon interactions are the extremely important research topics in the non-perturbation domain of quantum chromodynamics (QCD).
Since the discovery of the vector mesons, they have become attractive probes for the study of hadronic interaction.
Due to the behavior of the near threshold cross section is related to the vector meson-nucleon scattering length \cite{Gell-Mann:1961jim},
experimentally, the vector meson-nucleon interaction can be investigated using vector meson photoproduction within the Vector Meson Dominance (VMD) model.
The $\omega p$, $\phi p$, $J/\psi-p$, $\psi(2S)p$ and $\rho^{0}p$ scattering lengths have been fully analysed in Refs.
\cite{Strakovsky:2014wja,Strakovsky:2019bev,Strakovsky:2020uqs,Pentchev:2020kao,Wang:2022xpw,Wang:2022zwz}.
Based on the VMD model, the absolute value of the scattering length may be determined from the total near-threshold vector-meson photoproduction
cross-section \cite{Gell-Mann:1961jim} or the differential cross-section of vector-meson photoproduction at threshold \cite{Titov:2007xb}.

We try to extract the scattering length of the vector meson-neutron and vector meson-proton with current available experimental data.
In Ref. \cite{Strakovsky:2019bev}, they estimate the $J/\psi-p$ scattering length using the VMD model with the new GlueX data to the threshold.
Within the VMD model, the total $\gamma N\to$ $XN$ cross-section is related to both the total $XN \to XN$ cross-section at threshold energy and
the scattering length $|\alpha_{XN}|$ by \cite{Titov:2007xb}:
\begin{equation}
	\sigma^{\gamma N}\left(s_{t h r}\right)=\frac{\alpha \pi}{\gamma_{X}^{2}} \frac{q_{XN}}{k_{\gamma N}} \cdot \sigma^{XN}\left(s_{t h r}\right)=\frac{\alpha \pi}{\gamma_{X}^{2}} \frac{q_{XN}}{k_{\gamma N}} \cdot 4 \pi \alpha_{XN}^{2},
	\label{eq:total}
\end{equation}
where $\alpha$ = 1/137 denotes the fine structure constant,
and $X$ is a index which represents the vector mesons (e.g. $\omega$, $\phi$, $J/\psi$, etc.)
The $k_{\gamma N}$ and $q_{XN}$ in the above equation are the momenta in the center-of-mass of the initial and final state particles, respectively,
and $\gamma_{X}$ is the photon-vector meson coupling constant obtained from the $X \to e^+e^-$ decay width. 
The photon–$\omega$ and photon–$\phi$ coupling constants used in this work are 8.53 and 6.71 \cite{Workman:2022ynf}, respectively. 
Eq. (\ref{eq:total}) is taken at the threshold energy, where s$_{thr}$ = $(M + m)^{2}$ with M and m being the masses of the vector meson and nucleon, respectively.

To estimate the scattering length with the experimental measurements of the differential photoproduction cross-section, in this study,
we adopted the relation between the total cross-section and the differential cross-section at the threshold \cite{Pentchev:2020kao}.
The relationship between the scattering length and the differential cross-section is expressed as:
\begin{equation}
  \begin{split}
    \frac{d \sigma^{\gamma X}}{d t}\left(s_{t h r}, t=t_{thr}\right)=\frac{\alpha \pi}{\gamma_{X}^{2}} \frac{\pi}{k_{\gamma N}^{2}} \cdot \alpha_{ XN}^{2}. \\
    \frac{d \sigma^{\gamma X}}{d t}\left(s_{t h r}, t=0\right)=\frac{\alpha \pi}{\gamma_{X}^{2}} \frac{\pi}{k_{\gamma N}^{2}} \cdot \alpha_{ XN}^{2}.
    \label{eq:diffxsection}
  \end{split}
\end{equation}

A key problem in determining the scattering length at threshold t$_{thr}$ is to extrapolate the cross section to the point of  t$\rightarrow$t$_{thr}$ or s$\rightarrow$s$_{thr}$.
In addition, the $d\sigma^{\gamma X}/dt(s_{thr}, t=0)$ at left-hand side of Eq. (\ref{eq:diffxsection}) is not a directly measurable quantity,
as it requires extrapolation of the energy to the threshold and extrapolation of t from the physical region ($t_{min}<t<t_{max}$) to the non-physical point t = 0.
Therefore, when the vector meson nucleon scattering length is extracted from the differential cross-section data, we can not only extrapolate the energy at the threshold t$_{thr}$,
but also extrapolate t to t = 0.

In this work, the following exponential function was used to fit the differential cross-section data of near-threshold $\omega$ and $\phi$ photoproductions on deuterium target
or hydrogen target.
\begin{equation}
\begin{split}
\frac{d\sigma}{dt}=Ae^{-bt},
\end{split}
\label{eq:exp_fit}
\end{equation}
where $A=d\sigma/dt|_{t=0}$ denotes the forward differential cross-section and $b$ describes the slope parameter.
Combining Eq. (\ref{eq:diffxsection}) and Eq. (\ref{eq:exp_fit}), we can obtain the forward differential cross-section value
$d\sigma/dt|_{t=t_{thr}}$ and $d\sigma/dt|_{t=0}$ and extract the vector meson-nucleon scattering length $|\alpha_{XN}|$.

Great progress has been made in meson-proton scattering length measurements, but little is known about the vector meson-neutron scattering length.
As the vector meson-proton scattering length has already been fixed by the data of $\omega$, $\phi$ $J/\psi$, $\psi(2S)$ and $\rho^{0}$
vector meson photoproductions near threshold in the previous analyses \cite{Ishikawa:2019rvz,Titov:2007xb,Strakovsky:2019bev,Pentchev:2020kao,Wang:2022xpw,Wang:2022zwz},
We want to try whether we can extract the vector meson-neutron scattering length using the deuterium target data that the
neutron and the proton are loosely bound. The differential cross-section data of near-threshold $\omega$ and $\phi$ photoproductions on deuterium target were used for this analysis,
including the incoherent photoproduction of vector meson ($\omega$ and $\phi$ vector mesons) \cite{CBELSATAPS:2015wwn,LEPS:2009nuw} where the
neutron or the proton is knocked out.
The incoherent data of $\omega$ and $\phi$ photoproductions for scattering length extraction are provided by CBELSA/TAPS Collaboration and LEPS Collaboration, respectively.
In this work, we extract $\omega n$, $\omega p$ and $\phi$N scattering length from the differential cross-section data of near-threshold $\omega$
and $\phi$ photoproductions on deuterium target, where the deuteron is usually approximated as a weakly bound state of a quasi-free neutron plus a quasi-free proton.
For comparison study, we also extracted the scattering length $|\alpha_{\omega p}|$ of the free proton from the $\omega$ photoproduction on the hydrogen target by CBELSA/TAPS Collaboration.

\section{$\omega$$n$, $\omega$$p$ and $\phi$$N$ scattering lengths from $\omega$ and $\phi$ differential photoproduction cross sections}
\label{data_analysis_scattering_lengths}

\subsection{$\omega$n$^{*}$ scattering length from $\omega$ differential photoproduction cross sections on the deuterium target}
\label{omegan_scattering_lengths}

\begin{figure}[htp]
\centering
\includegraphics[width=0.5\textwidth]{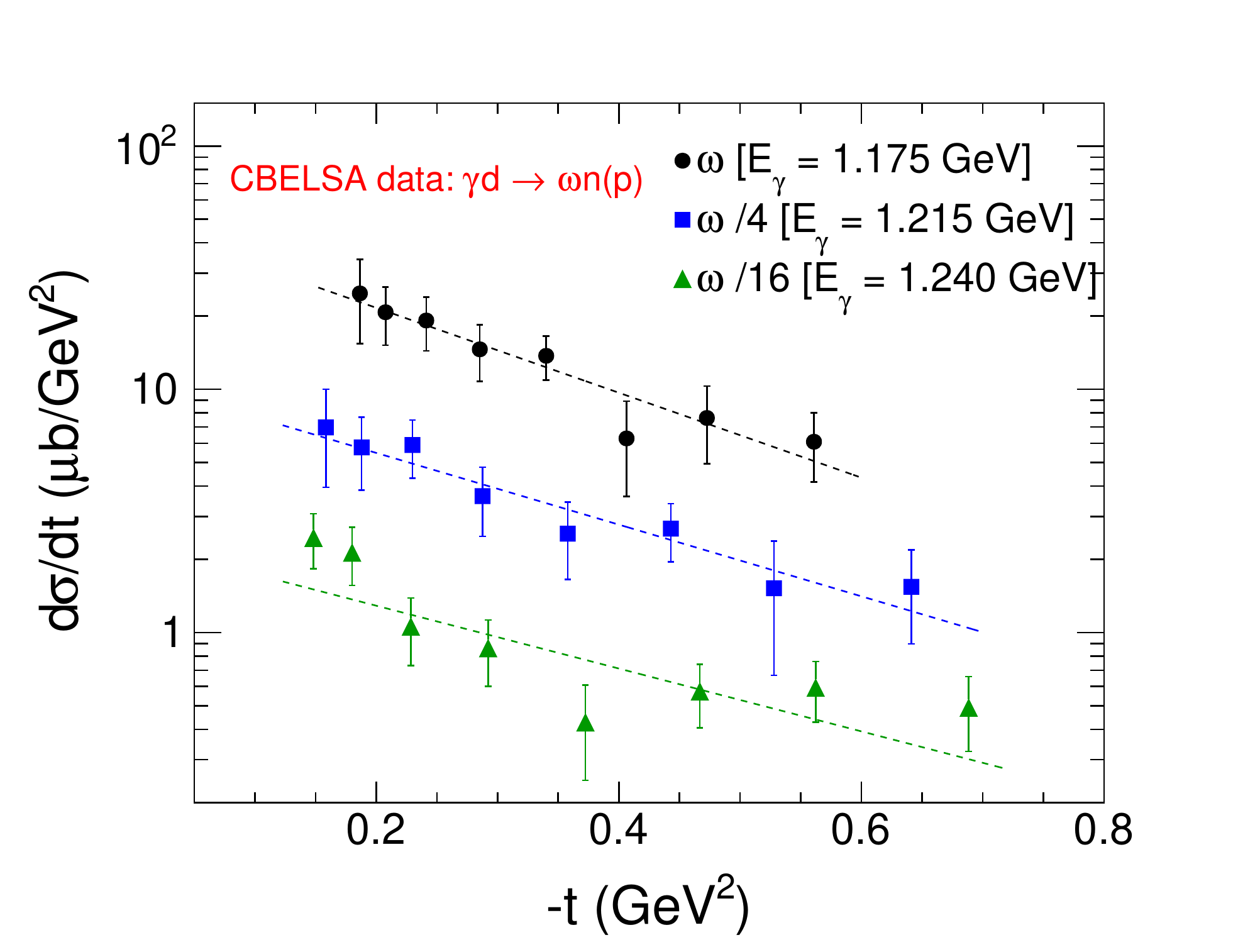}
\caption{
Differential cross sections of the near threshold photoproduction of $\omega$ meson produced off the quasi-free neutron versus the momentum
transfer -t in the deuterium target \cite{CBELSATAPS:2015wwn}.
The three incident photon energies $E_{\gamma}$ (1.175, 1.215 and 1.240 GeV) near the threshold of $\omega$ mesons produced off the quasi-free neutron
are marked in the figure.
Some cross sections are scaled using the coefficients shown in the figure.
}
\label{fig:Omega_neutron_sl}
\end{figure}

\begin{table*}[h]
  \caption{
    The absolute value of the extracted $\omega$n$^{*}$ scattering length obtained from the differential cross-section data of $\omega$ meson produced off
    the bound neutron in deuteron near threshold at different photon incident energies using different extrapolating methods.
  }
  \begin{center}
    \begin{ruledtabular}
      \begin{tabular}{ cccc }
        $E_{\gamma}$ (GeV)                                                    &      1.175         &      1.215          &  1.240        \\
        \hline
        $|\alpha_{\omega n^{*}}|$ (fm) ($d\sigma^{\gamma n^{*}}/dt(s_{thr},t_{thr})$)    &  $0.683\pm 0.163$  &  $0.717\pm 0.125$   &  $0.715\pm 0.124$    \\
        $|\alpha_{\omega n^{*}}|$ (fm) ($d\sigma^{\gamma n^{*}}/dt(s_{thr},0)$)          &  $1.337\pm 0.226$  &  $1.263\pm 0.224$   &  $1.176\pm 0.222$    \\
      \end{tabular}
    \end{ruledtabular}
  \end{center}
  \label{tab:wn_List}
\end{table*}
Figure \ref{fig:Omega_neutron_sl} shows the differential cross sections of the $\omega$ meson photoproductions
produced off the bound neutron as a function of momentum transfer $-t$, with three incident photon energies E$_{\gamma}$.
This reaction with exactly four neutral hits ($\omega$ $\rightarrow$ $\gamma\gamma\gamma$ and neutron)
of the experiment at ELSA \cite{CBELSATAPS:2015wwn} is detected where the $\omega$ meson was produced off the bound neutron
in the liquid deuterium target giving the quasi-free reaction $\gamma$d $\rightarrow$ $\omega$n(p).
The differential cross sections of the $\left|t \right|$-dependence are fitted with an exponential function.
We determined the parameters A and the b at three incident photon energies
$E_{\gamma}$ (1.175, 1.215 and 1.240 GeV) through fitting to the $\omega$ mesons produced off the bound neutron with
deuterium target data \cite{CBELSATAPS:2015wwn} with Eq. (\ref{eq:exp_fit}).
Then the values of $d\sigma^{\gamma n^{*}}/dt(s_{thr},t_{thr})$ and $d\sigma^{\gamma n^{*}}/dt(s_{thr},0)$ are calculated
by fitting parameters A and b obtained by fitting the $\left|t \right|$-dependent differential photoproduction cross-section data.
The scattering length $|\alpha_{\omega n^{*}}|$ at the threshold t$_{thr}$ and in t to t = 0 are then calculated by Eq. (\ref{eq:diffxsection}).
The extracted scattering length $|\alpha_{\omega n^{*}}|$ are listed in Table \ref{tab:wn_List}.
The averaged scattering length $|\alpha_{\omega n^{*}}|$ for the three extracted values at different E$_{\gamma}$ energies at threshold
is calculated to be 0.709 $\pm$ 0.078 fm.
The averaged scattering length $|\alpha_{\omega n^{*}}|$ for the three extracted values at different E$_{\gamma}$ energies in t to t = 0
is calculated to be 1.258 $\pm$ 0.130 fm.
Here we use the following formula to calculate the weighted average: $\overline{x}$ $\pm$ $\delta$ $\overline{x}$ = $\Sigma_{i}$$\omega_{i}$$x_{i}$/$\Sigma_{i}$$\omega_{i}$$\pm$
($\Sigma_{i}$$\omega_{i}$)$^{-1/2}$ with $\omega_{i}$ = 1/($\delta x_{i}$)$^{2}$ \cite{ParticleDataGroup:2020ssz}.
It is worth noting that the weighted average values of scattering length at different energies obtained here are consistent with the simultaneous fitting results of all data sets.

\subsection{$\omega$p$^{*}$ scattering length from $\omega$ differential photoproduction cross sections on the deuterium target}
\label{omegap_scattering_lengths}

\begin{figure}[htp]
\centering
\includegraphics[width=0.5\textwidth]{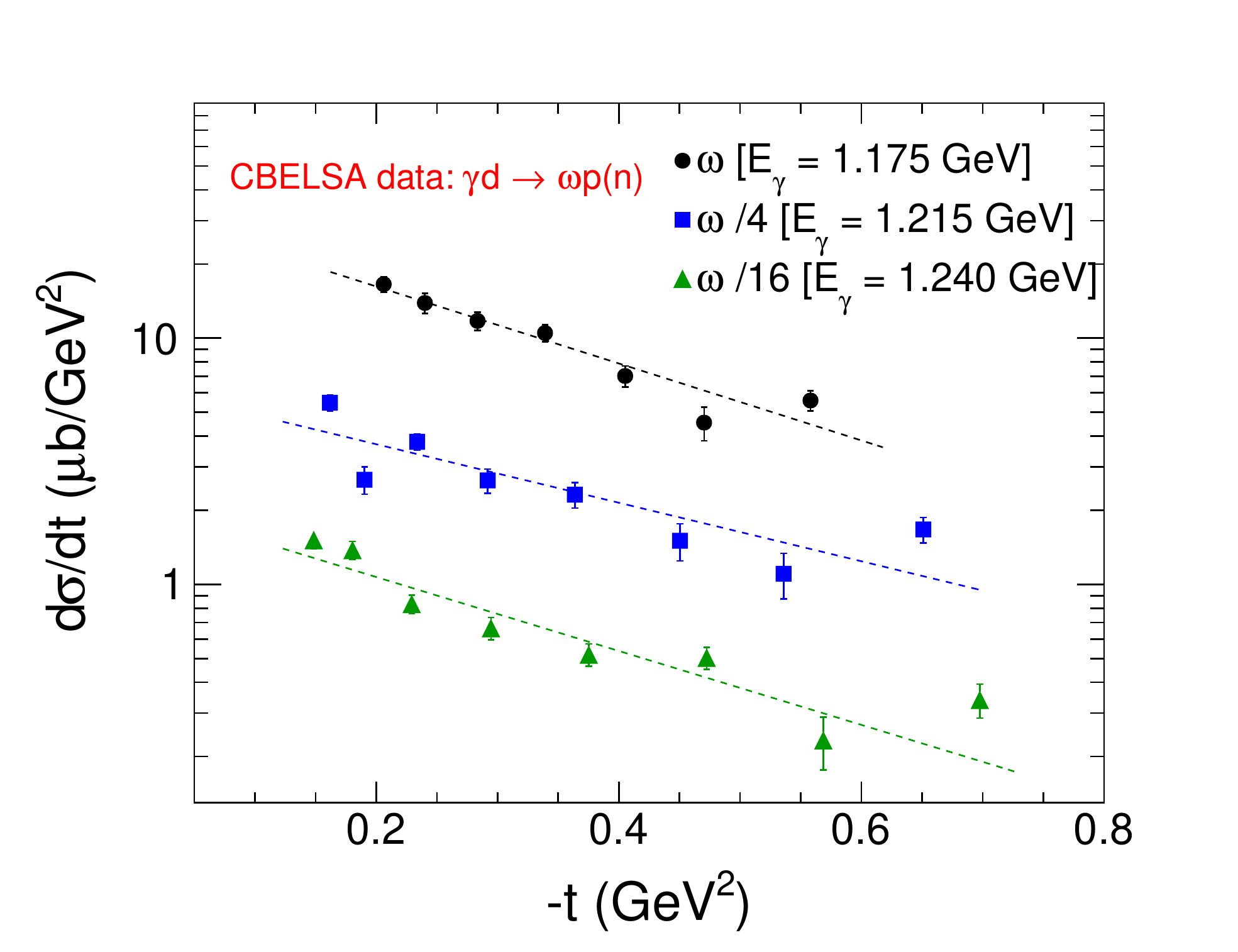}
\caption{
  Differential cross sections of the photoproduction of $\omega$ meson produced off the quasi-free proton versus the momentum
  transfer -t in deuterium near the thresholds \cite{CBELSATAPS:2015wwn}.
  The three incident photon energies $E_{\gamma}$ (1.175, 1.215 and 1.240 GeV) near the threshold of $\omega$ meson produced off the quasi-free proton are marked in the figure.
  Some cross sections are scaled using the coefficients shown in the figure.
}
\label{fig:Omega_proton_sl}
\end{figure}
\begin{table*}[h]
  \caption{
    The absolute value of the extracted $\omega$p$^{*}$ scattering length obtained from the differential cross-section data of $\omega$ meson produced off the bound proton in deuteron
    near threshold at different photon incident energies using different extrapolating methods.
  }
  \begin{center}
    \begin{ruledtabular}
      \begin{tabular}{ cccc }
        $E_{\gamma}$ (GeV)                       &      1.175          &      1.215          &  1.240        \\
        \hline
        $|\alpha_{\omega p^{*}}|$ (fm) ($d\sigma^{\gamma p^{*}}/dt(s_{thr},t_{thr})$)    &  $0.608\pm 0.048$  &  $0.616\pm 0.033$   &  $0.631\pm 0.035$    \\
        $|\alpha_{\omega p^{*}}|$ (fm) ($d\sigma^{\gamma p^{*}}/dt(s_{thr},0)$)          &  $1.110\pm 0.063$  &  $0.973\pm 0.049$   &  $1.125\pm 0.056$    \\
      \end{tabular}
    \end{ruledtabular}
  \end{center}
  \label{tab:wp_List}
\end{table*}

Figure \ref{fig:Omega_proton_sl} shows the differential cross sections of the $\omega$ meson
photoproductions produced off the bound proton as a function of momentum transfer $-t$.
This reaction with exactly three neutral hits ($\omega$ $\rightarrow$ $\gamma\gamma\gamma$) and one charged hit (proton)
of the experiment at ELSA \cite{CBELSATAPS:2015wwn} is detected where the $\omega$ meson was produced off the bound proton in the liquid deuterium target
giving the quasi-free reaction $\gamma$d $\rightarrow$ $\omega$p(n).
The differential cross sections of the $\left|t \right|$-dependence are fitted with an exponential function.
We determined the parameters A and the b at three incident photon energies $E_{\gamma}$ (1.175, 1.215 and 1.240 GeV)
by fitting the $\omega$ mesons produced off the bound proton on the deuterium target data \cite{CBELSATAPS:2015wwn} with Eq. (\ref{eq:exp_fit}).
The values of $d\sigma^{\gamma p^{*}}/dt(s_{thr},t_{thr})$ and $d\sigma^{\gamma p^{*}}/dt(s_{thr},0)$ are calculated by fitting parameters A and b
obtained by fitting the $\left|t \right|$-dependent differential photoproduction cross-section data.
The scattering length $|\alpha_{\omega p^{*}}|$ at the threshold t$_{thr}$ and in t to t=0 are then calculated by Eq. (\ref{eq:diffxsection}).
The extracted scattering length $|\alpha_{\omega p^{*}}|$ are listed in Table \ref{tab:wp_List}.
The averaged scattering length $|\alpha_{\omega p^{*}}|$ for the three extracted values at different E$_{\gamma}$ energies at threshold
is calculated to be 0.621 $\pm$ 0.022 fm.
The averaged scattering length $|\alpha_{\omega p^{*}}|$ for the three extracted values at different E$_{\gamma}$ energies in t to t = 0
is calculated to be 1.056 $\pm$ 0.032 fm.

\subsection{$\omega$p scattering length from $\omega$ differential photoproduction cross sections on the hydrogen target}
\label{omegap_scattering_lengths_hydrogen_target}

\begin{figure}[htp]
\centering
\includegraphics[width=0.5\textwidth]{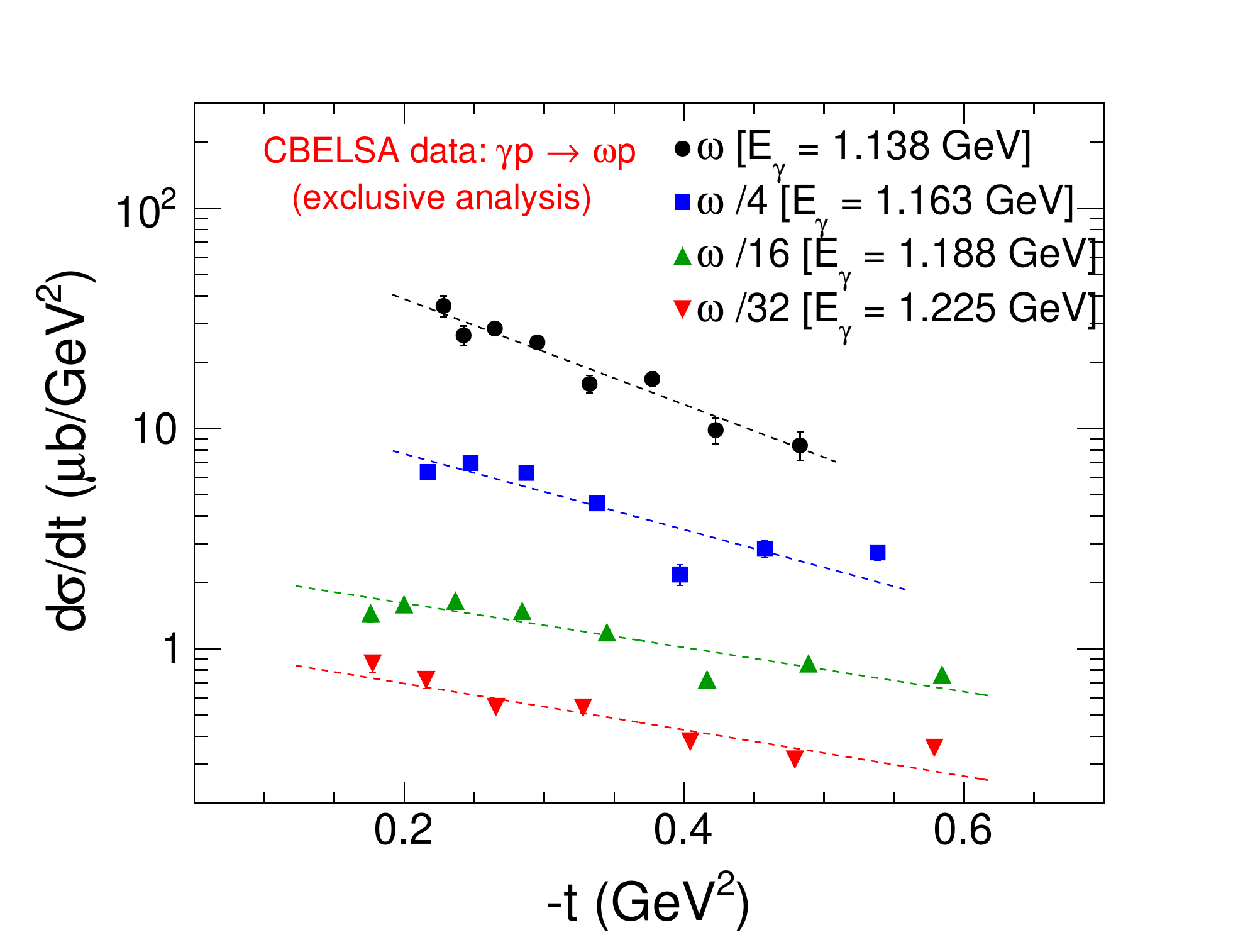}
\caption{
  Differential cross sections of the photoproductions of $\omega$ meson produced off the free proton near the threshold
  versus the momentum transfer -t in the hydrogen target \cite{CBELSATAPS:2015wwn}.
  The four incident photon energies $E_{\gamma}$ (1.138, 1.163, 1.188 and 1.225 GeV) near the threshold of $\omega$ meson are marked in the figure.
  The exclusive experimental data results by CBELSA/TAPS Collaboration are used for the extraction of the $\omega p$ scattering length.
  Some cross sections are scaled using the coefficients shown in the figure.
}
\label{fig:Omega_FreeProton_sl_Exclusive}
\end{figure}
\begin{table*}[h]
  \caption{
    The absolute value of the extracted $\omega$p scattering length obtained from the differential cross-section data of $\omega$ meson produced off the free proton
    near threshold with exclusive analysis at different photon energies using different extrapolating methods.
  }
  \begin{center}
    \begin{ruledtabular}
      \begin{tabular}{ ccccccccc }
        $E_{\gamma}$ (GeV)                       &      1.138         &      1.163         &  1.188              & 1.225\\
        \hline
        $|\alpha_{\omega p}|$ (fm) ($d\sigma^{\gamma p}/dt(s_{thr},t_{thr})$)    &  $0.827\pm 0.091$  &  $0.816\pm 0.043$   &  $0.835\pm 0.029$ &  $0.770\pm 0.034$    \\
        $|\alpha_{\omega p}|$ (fm) ($d\sigma^{\gamma p}/dt(s_{thr},0)$)          &  $2.077\pm 0.158$  &  $1.581\pm 0.081$   &  $1.229\pm 0.044$ &  $1.155\pm 0.057$    \\
      \end{tabular}
    \end{ruledtabular}
  \end{center}
  \label{tab:FreeProtonslList_Omega_Exclusive}
\end{table*}

\begin{figure}[htp]
\centering
\includegraphics[width=0.5\textwidth]{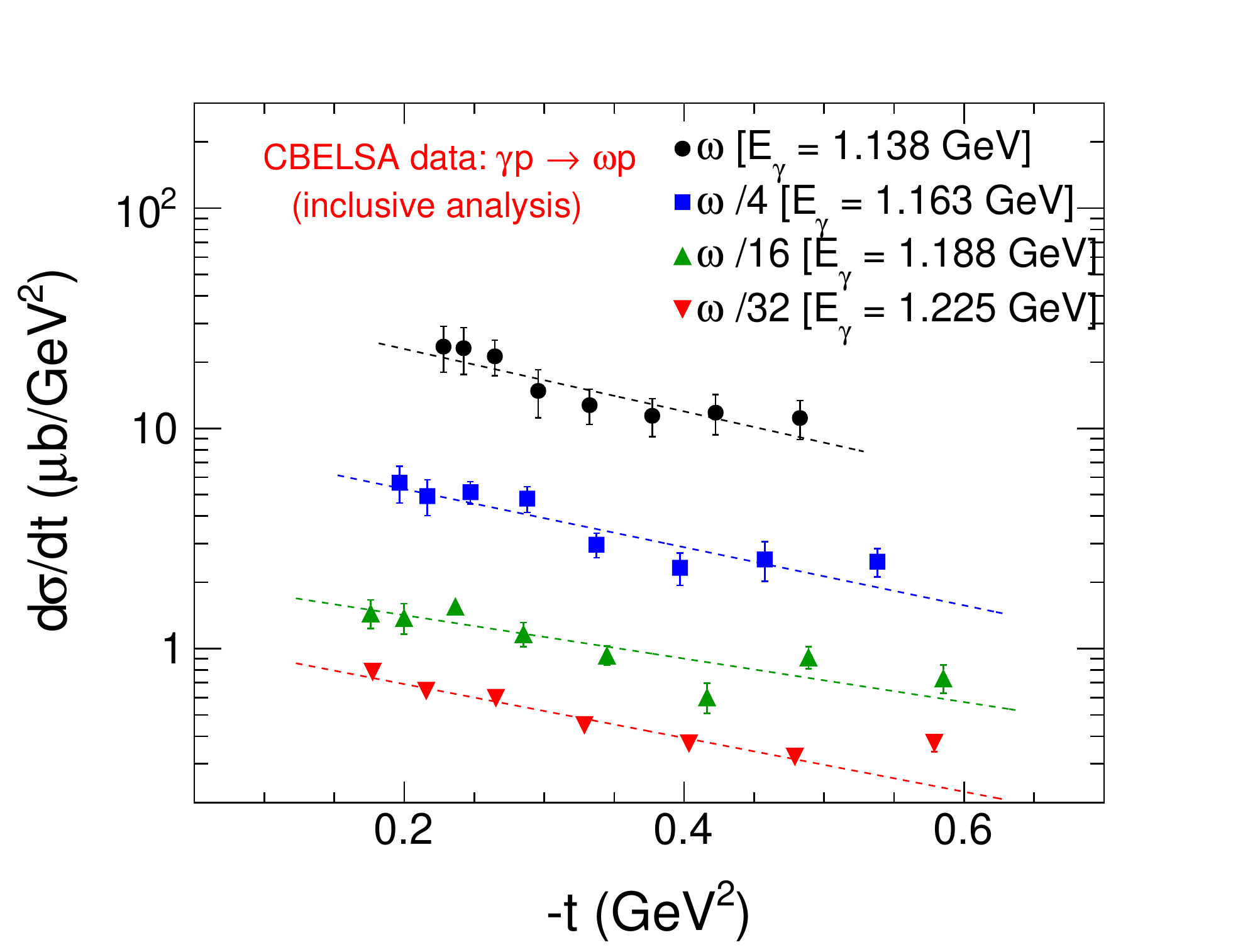}
\caption{
  Differential cross sections of the photoproduction of $\omega$ meson produced off the free proton near the threshold
  versus the momentum transfer -t in the hydrogen target \cite{CBELSATAPS:2015wwn}.
  The four incident photon energies $E_{\gamma}$ (1.138, 1.163, 1.188 and 1.225 GeV) near the threshold of $\omega$ meson are marked in the figure.
  The inclusive experimental data by CBELSA/TAPS Collaboration are used for the extraction of the $\omega p$ scattering length.
  Some cross sections are scaled using the coefficients shown in the figure.
}
\label{fig:Omega_FreeProton_sl_Inclusive}
\end{figure}
\begin{table*}[h]
  \caption{
    The absolute value of the extracted $\omega$p scattering length obtained 
    from the differential cross-section data of $\omega$ meson produced off the free proton
    near threshold with inclusive analysis at different photon energies using different extrapolating methods.
  }
  \begin{center}
    \begin{ruledtabular}
      \begin{tabular}{ cccccccccc }
        $E_{\gamma}$ (GeV)                  &      1.138         &      1.163         &  1.188              & 1.225\\
        \hline
        $|\alpha_{\omega p}|$ (fm) ($d\sigma^{\gamma p}/dt(s_{thr},t_{thr})$)    &  $0.740\pm 0.192$  &  $0.722\pm 0.076$   &  $0.787\pm 0.054$ &  $0.748\pm 0.024$    \\
        $|\alpha_{\omega p}|$ (fm) ($d\sigma^{\gamma p}/dt(s_{thr},0)$)          &  $1.277\pm 0.237$  &  $1.200\pm 0.125$   &  $1.150\pm 0.081$ &  $1.197\pm 0.033$    \\
      \end{tabular}
    \end{ruledtabular}
  \end{center}
  \label{tab:FreeProtonslList_Omega_Inclusive}
\end{table*}

Figure \ref{fig:Omega_FreeProton_sl_Exclusive} and Figure \ref{fig:Omega_FreeProton_sl_Inclusive} show the differential cross sections of the $\omega$ meson
photoproductions produced off the free proton on the liquid hydrogen target as a function of momentum transfer $-t$ of the exclusive analysis and the
inclusive analysis \cite{CBELSATAPS:2015wwn}, respectively.
In the exclusive analysis, the recoil nucleon was identified in coincidence.
In the inclusive analysis, there is no condition for detection of the recoil nucleon. 
We determined the parameters A and the b at four incident photon energies $E_{\gamma}$ (1.175, 1.215 and 1.240 GeV) by fitting the $\omega$ mesons
produced off the free proton in liquid hydrogen target with Eq. (\ref{eq:exp_fit}).
The values of $d\sigma^{\gamma p}/dt(s_{thr},t_{thr})$ and $d\sigma^{\gamma p}/dt(s_{thr},0)$ are calculated
by fitting parameters A and b obtained by fitting the $\left|t \right|$-dependent differential photoproduction cross-section data.
The scattering length $|\alpha_{\omega p}|$ at the threshold t$_{thr}$ and in t to t = 0 are then calculated by Eq. (\ref{eq:diffxsection}).
The extracted scattering length $|\alpha_{\omega p}|$ of the exclusive analysis and the inclusive analysis
are listed in Table \ref{tab:FreeProtonslList_Omega_Exclusive} and \ref{tab:FreeProtonslList_Omega_Inclusive}, respectively.
The averaged scattering lengths $|\alpha_{\omega p}|$ of the exclusive analysis and inclusive analysis at different E$_{\gamma}$ energies at threshold
are calculated to be 0.811 $\pm$ 0.019 fm and 0.752 $\pm$ 0.021 fm, respectively.
The averaged scattering length $|\alpha_{\omega p}|$ of the exclusive analysis and inclusive analysis at different E$_{\gamma}$ energies in t to t = 0 
are calculated to be 1.293 $\pm$ 0.032 fm and 1.192 $\pm$ 0.030 fm, respectively.
We see that the $\omega$p scattering length of the free proton based on the exclusive data and the inclusive data agree with each other.
We also find that the free proton $\omega$p scattering length may be larger than the bound proton in the deuteron, but more precise data are needed to further check this.
The A2 Collaboration at MAMI conducted an experimental of $\omega$ photoproduction on the proton and estimated that the $\omega$p scattering length
was equal to 0.82 $\pm$ 0.03 fm \cite{Strakovsky:2014wja} based on the VMD model,
which is consistent with the result of our extraction through incoherent experimental data of $\omega$ photoproductions provided by
CBELSA/TAPS Collaboration \cite{CBELSATAPS:2015wwn}.
The obtained scattering length $|\alpha_{\omega p}|$ in this work are consistent with the experimental measurements of photoproduction
of the $\omega$ meson on the proton near the threshold by Ishikawa $|\alpha_{\omega p}|$ = 0.97 $\pm$ 0.16 
fm \cite{Ishikawa:2019rvz}.
In short, our results of the $\omega$p scattering length are consistent with previous experimental measurements and extration results based on some theoretical models.

\subsection{$\phi$N scattering length from $\phi$ differential photoproduction cross sections on the deuterium target}
\label{phip_scattering_lengths}

\begin{figure}[htp]
\centering
\includegraphics[width=0.50\textwidth]{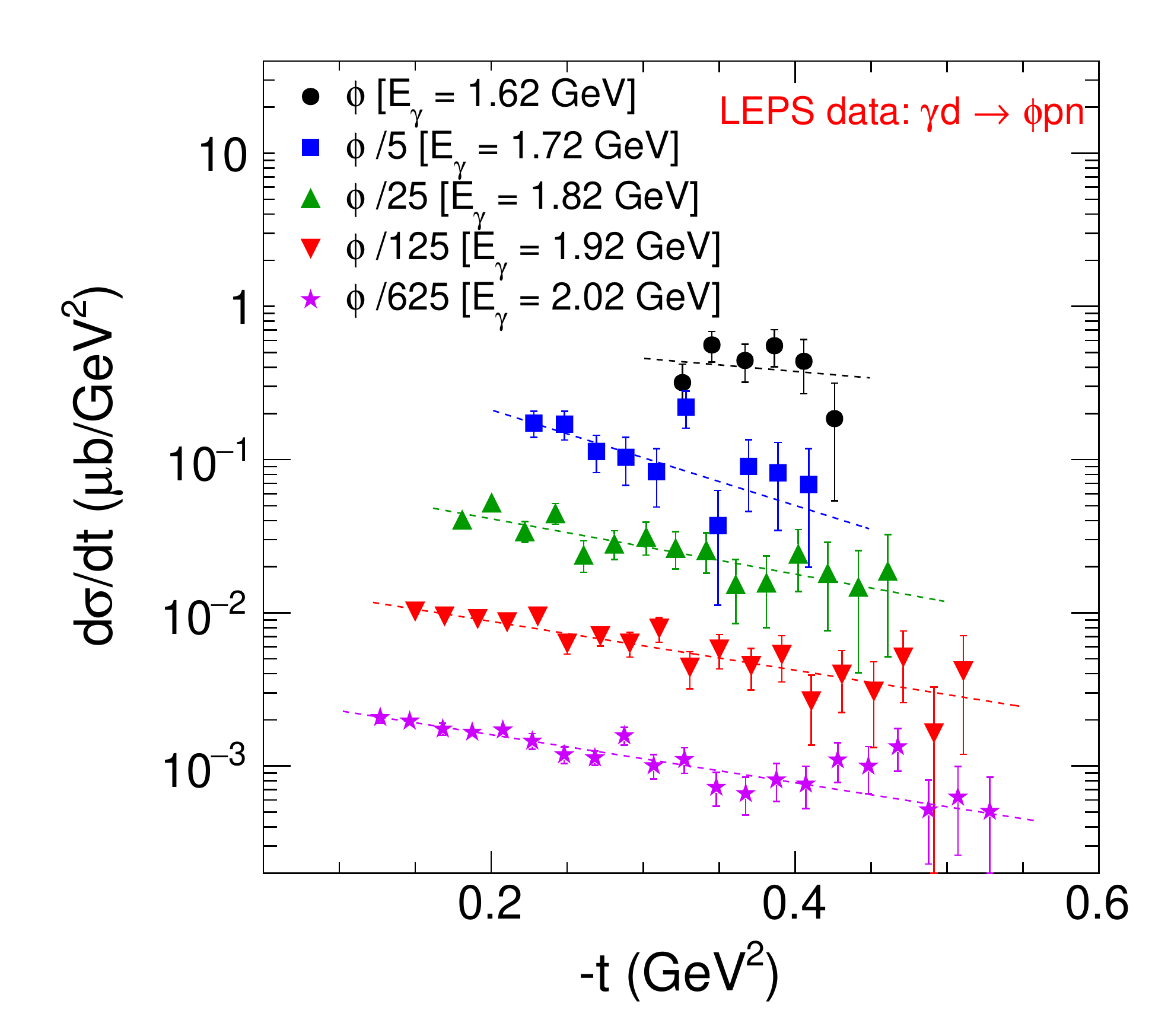}
\caption{
  Differential cross sections of the incoherent $\phi$ meson photoproduction ($\gamma$d $\rightarrow$ $\phi$pn)
  near the threshold as a function of the momentum transfer -t off the quasi-free nucleon in the deuterium target \cite{LEPS:2009nuw}.
  The five incident photon energies $E_{\gamma}$ (1.62, 1.72, 1.82, 1.92 and 2.02 GeV) near the threshold of incoherent $\phi$ meson photoproduction are marked in the figure.
  Some cross sections are scaled using the coefficients shown in the figure.
}
\label{fig:Phi_nucleon_Rm}
\end{figure}

\begin{table*}
  \caption{
    The absolute value of the extracted $\phi$N scattering length obtained from
    from the differential cross sections of incoherent $\phi$ mesons photonproduction from the deuteron
    near threshold at different photon energies using different extrapolating methods.
  }
  \begin{center}
    \begin{ruledtabular}
      \begin{tabular}{ cccccccccc }
        $E_{\gamma}$ (GeV)                &      1.62         &      1.72         &  1.82              & 1.92                  & 2.02\\
        \hline
        $|\alpha_{\phi N}|$ (fm) ($d\sigma^{\gamma N}/dt(s_{thr},t_{thr})$)    &  $0.105\pm 0.114$  &  $0.066\pm 0.041$   &  $0.103\pm 0.024$ &  $0.114\pm 0.015$ &  $0.110\pm 0.011$  \\
        $|\alpha_{\phi N}|$ (fm) ($d\sigma^{\gamma N}/dt(s_{thr},0)$)          &  $0.171\pm 0.110$  &  $0.397\pm 0.135$   &  $0.290\pm 0.034$ &  $0.285\pm 0.018$ &  $0.271\pm 0.012$   \\
      \end{tabular}
    \end{ruledtabular}
  \end{center}
  \label{tab:RadiusListPhiMeson}
\end{table*}

In this following analysis, the scattering length $|\alpha_{\phi N}|$ refers to the averaged value
of the $|\alpha_{\omega n^{*}}|$ and $|\alpha_{\omega p^{*}}|$ of quasi-free neutron and proton in the deuteron.
Figure \ref{fig:Phi_nucleon_Rm} shows the differential cross sections of incoherent $\phi$ meson
photoproductions from the deuteron as a function of momentum transfer $-t$.
For the incoherent $\phi$ meson photoproduction from the deuteron in the LEPS spectrometer \cite{LEPS:2009nuw},
a cut for the $\phi$ meson selection is performed on the missing mass spectra, to ensure the $\phi$ meson
was interacting with the individual nucleon rather than the whole deuteron.
The differential cross sections of the $\left|t \right|$-dependence are still fitted with an exponential function.
We determined the parameters A and the b with five incident photon energies
$E_{\gamma}$ (1.62, 1.72, 1.82, 1.92 and 2.02 GeV) by fitting the $\phi$ meson produced off the quasi-free proton or quasi-free neutron
in the deuterium target with Eq. (\ref{eq:exp_fit}).
The value of $d\sigma^{\gamma N}/dt(s_{thr},t_{thr})$ and $d\sigma^{\gamma N}/dt(s_{thr},0)$ are calculated
by fitting parameters A and b obtained by fitting the $\left|t \right|$-dependent differential photoproduction cross-section data.
The scattering length $|\alpha_{\phi N}|$ at the threshold t$_{thr}$ and in t to t = 0 are then calculated by Eq. (\ref{eq:diffxsection}).
The extracted scattering length $|\alpha_{\phi N}|$ are listed in Table \ref{tab:RadiusListPhiMeson}.
The averaged scattering length $|\alpha_{\phi N}|$ of incoherent $\phi$ meson photoproduction in deuteron at different E$_{\gamma}$ energies at threshold is calculated to be
0.109 $\pm$ 0.008 fm at threshold t$_{thr}$ and 0.276 $\pm$ 0.010 fm in t to t = 0, respectively,
The scattering length $|\alpha_{\phi N}|$ at threshold t$_{thr}$ obtained in this work is a little larger than the estimated value $\phi$p at threshold t$_{thr}$ extracted from the $\phi$-meson photoproduction data from the CLAS experiment in Hall B of Jefferson Laboratory \cite{Strakovsky:2020uqs}.
Furthermore, the value for $|\alpha_{\phi N}|$ at threshold t$_{thr}$ as determined in this paper is smaller than
the $|\alpha_{\phi N}|$ result 0.15 fm given from forward coherent $\phi$-meson photoproduction from deuterons near threshold by the LEPS Collaboration \cite{LEPS:2005hax,Chang:2007fc},
the result 0.15 $\pm$ 0.02 fm using a new QCD sum rule analysis on the spin-isospin averaged $\phi$ meson-nucleon scattering \cite{Koike:1996ga}
and the estimation value with $\phi$N potential approaches \cite{Gao:2000az,Titov:2007xb}.

\begin{figure}[htp]
\centering
\includegraphics[width=0.5\textwidth]{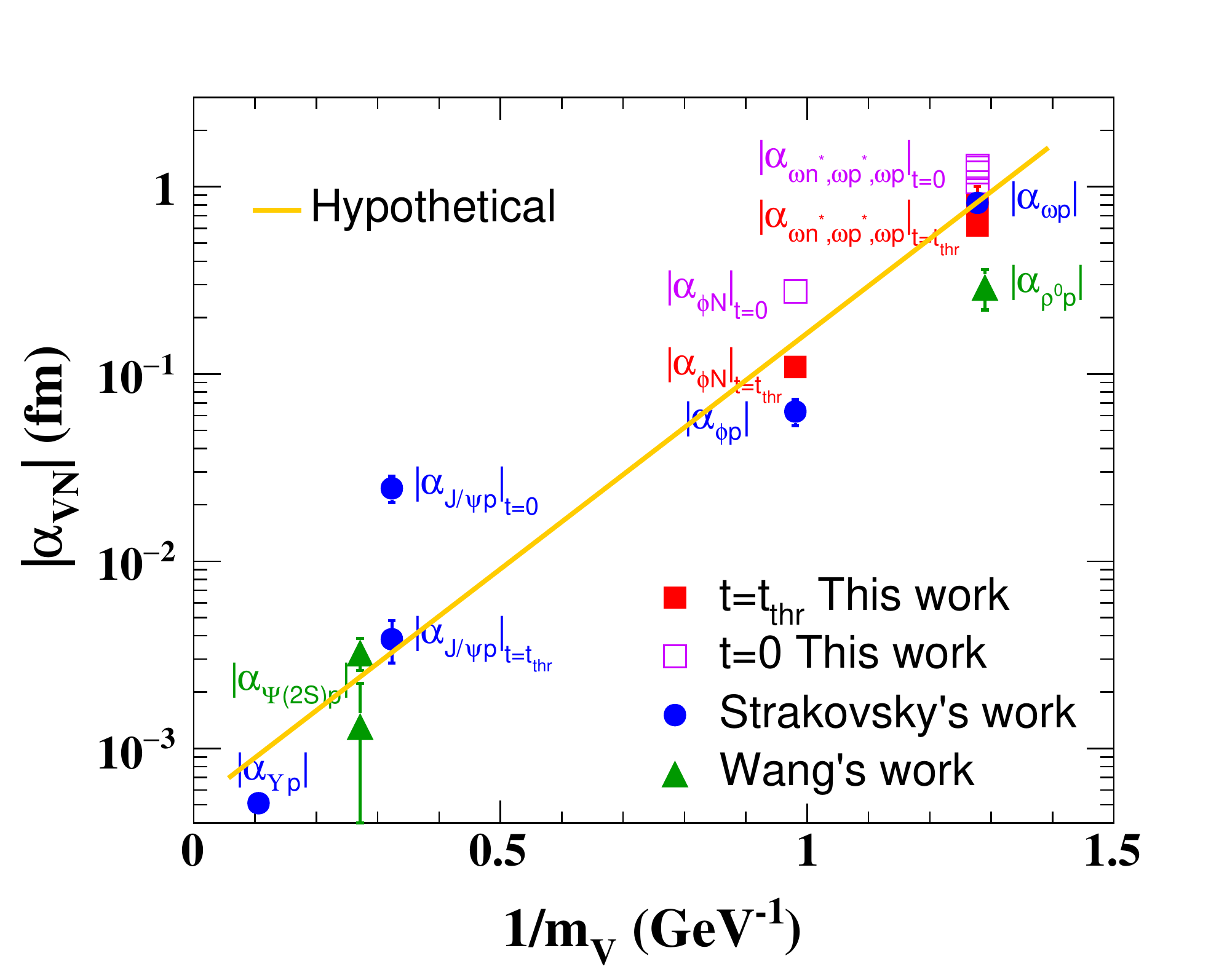}
\caption{
  Comparison of the scattering lengths $|\alpha_{VN}|$ as a function of the inverse mass of vector mesons,
  including $\omega$, $\rho^{0}$, $\phi$, $J/\psi$, $\psi(2S)$, and $\Upsilon$.
  The red squares show the $|\alpha_{\omega n^{*}}|$, $|\alpha_{\omega p^{*}}|$, $|\alpha_{\omega p}|$ and $|\alpha_{\phi N}|$ scattering length at threshold t$_{thr}$ from this work.
  The violet hollow squares show the $|\alpha_{\omega n^{*}}|$, $|\alpha_{\omega p^{*}}|$, $|\alpha_{\omega p}|$ and $|\alpha_{\phi N}|$ scattering length in t to t = 0 from this work.
  The blue cricles show the Strakovsky's analysis of $\omega$ \cite{Strakovsky:2014wja}, $\phi$ \cite{Strakovsky:2020uqs}, $J/\psi$ \cite{Pentchev:2020kao}, and $\Upsilon$ \cite{Strakovsky:2021vyk}.
  The green triangles shows the Wang's analysis $\rho^{0}$ \cite{Wang:2022zwz} and $\psi(2S)$ \cite{Wang:2022xpw}.
  The orange solid line is hypothetical \cite{Strakovsky:2021vyk}.
}
\label{fig:avN_1mv}
\end{figure}

\section{Discussion and Summary}
\label{Discussion and Summary}

\begin{table*}[h]
  \caption{
    The ratio of the scattering length $|\alpha_{\omega n^{*}}|$ to the scattering length $|\alpha_{\omega p^{*}}|$ with the deuterium target data and
    the ratio of the scattering length $|\alpha_{\omega p}|$ of the hydrogen target data from the exclusive analysis and from the inclusive analysis
    to the scattering length $|\alpha_{\omega p^{*}}|$ with the deuterium target data.
  }
  \begin{center}
    \begin{ruledtabular}
      \begin{tabular}{ cccc }
        $Category$                            &  $|\alpha_{\omega n^{*}}|$/$|\alpha_{\omega p^{*}}|$  &  $|\alpha_{\omega p}|$/$|\alpha_{\omega p^{*}}|$  \\
        \hline
        Ratio ($d\sigma^{\gamma p}/dt(s_{thr},t_{thr})$)  &  $1.142\pm 0.116$                   &  $1.263\pm 0.040$ \\
        Ratio ($d\sigma^{\gamma p}/dt(s_{thr},0)$)        &  $1.191\pm 0.108$                   &  $1.174\pm 0.035$ \\
        Average Ratio                                     &  $1.168\pm 0.079$                   &  $1.213\pm 0.026$ \\
        
      \end{tabular}
    \end{ruledtabular}
  \end{center}
  \label{tab:RadiusRatioList}
\end{table*}


Based on the VMD model and assuming energy independence of the differential cross-section, we analyzed the differential cross-section                                                 
experimental data of $\omega$ and $\phi$ photoproductions on deuterium and hydrogen targets and extracted the scattering lengths of $\omega$n, $\omega$p and $\phi$N.
In this study, we extrapolated the scattering lengths of $\omega$n, $\omega$p and $\phi$N not only to the energy at threshold t$_{thr}$, but also in t to t = 0 when using
the differential cross section data.
The $\omega$n$^{*}$ scattering lengths 0.709 $\pm$ 0.078 fm at threshold t$_{thr}$ and 1.258 $\pm$ 0.13 fm in t to t = 0 are obtained
for the first time from deuterium target data of $\omega$ photoproductions off the quasi-free neutron near the threshold energies.
The scattering length $|\alpha_{\omega p^{*}}|$ of the bound proton inside the deuteron from the $\omega$ photoproduction on the deuterium target
not only to the energy at threshold t$_{thr}$, but also in t to t = 0.
For a comparison study, we also extracted scattering length $|\alpha_{\omega p}|$ of the free proton from the $\omega$ photoproduction on the hydrogen target by CBELSA/TAPS Collaboration
at threshold t$_{thr}$ and in t to t = 0.
The extracted $\omega$p scattering length in this work are consistent with the experimental measurements of photoproduction
of the $\omega$ meson on the proton near the threshold \cite{CBELSATAPS:2015wwn, Ishikawa:2019rvz}.
Furthermore, we determined the scattering length $|\alpha_{\phi N}|$ with deuterium target data to be 0.109 $\pm$ 0.008 fm at threshold and 0.276 $\pm$ 0.010 fm in t to t = 0
from the incoherent $\phi$ photoproduction data on the deuterium target.
The obtained scattering length $|\alpha_{\phi N}|$ is in qualitative agreement with the experimental indications \cite{Strakovsky:2020uqs, LEPS:2005hax, Chang:2007fc} and
the theoretical predictions \cite{Koike:1996ga,Gao:2000az,Titov:2007xb}.
The above results should provide useful theoretical information for an in-depth understanding of nucleon interaction with vector mesons.

In order to compare and find the relationship between the scattering length $|\alpha_{VN}|$ and the vector meson mass, we have summarized
the results of different vector meson ($\omega$, $\rho^{0}$, $\phi$, $J/\psi$, $\psi(2S)$ and $\Upsilon$) scattering lengths previously
extracted by Strakovsky \cite{Strakovsky:2014wja, Strakovsky:2020uqs, Pentchev:2020kao, Strakovsky:2021vyk} and Wang \cite{Wang:2022zwz,Wang:2022xpw}
based on different extrapolation models and datasets and the results of scattering lengths $|\alpha_{\omega n^{*}}|$, $|\alpha_{\omega p^{*}}|$,
$|\alpha_{\omega p}|$ and $|\alpha_{\phi N}|$ extracted in this work.
Figure \ref{fig:avN_1mv} shows the comparison of the scattering lengths $|\alpha_{VN}|$ as a function of the inverse mass of vector mesons,
including $\omega$, $\rho^{0}$, $\phi$, $J/\psi$, $\psi(2S)$, and $\Upsilon$.
The scattering length $|\alpha_{VN}|$ is approximately exponential with the inverse mass of vector mesons.

In Table \ref{tab:RadiusRatioList}, we list the ratio of the scattering length $|\alpha_{\omega n^{*}}|$ to the scattering length $|\alpha_{\omega p^{*}}|$ with the deuterium target data and
the ratio of the scattering length $|\alpha_{\omega p}|$ of the hydrogen target data from the exclusive analysis and from the inclusive analysis
to the scattering length $|\alpha_{\omega p^{*}}|$ with the deuterium target data.
By comparing the extraction results, we find that the obtained $|\alpha_{\omega n^{*}}|$ of quasi-free neutron inside the deuteron is about 16.8 $\pm$ 7.9$\%$ larger than
the obtained $|\alpha_{\omega p^{*}}|$ of quasi-free proton inside the deuteron.
The obtained $|\alpha_{\omega p^{*}}|$ of bound proton in deuteron is about 21.3 $\pm$ 2.6$\%$ smaller than the $\omega$p scattering length of free proton.
This result indicates that the scattering length $|\alpha_{\omega p}|$ of the free proton may be larger than that of the bound proton inside deuteron.

To obtain the scattering length $|\alpha_{\omega n^{*}}|$ of the nearly free neutron, the tagging-spectator technique used
in BONuS \cite{CLAS:2011qvj,CLAS:2014jvt} and ALERT experiment \cite{Armstrong:2017zcm, Armstrong:2017wfw, Armstrong:2017zqr} is expected.
Therefore, we suggest a future experiment of the near-threshold $\omega$, $\rho^{0}$, $\phi$, $J/\psi$ and $\psi(2S)$ photoproduction on the deuterium target
with the low-momentum spectator proton tagged, to better understand the scattering lengths difference between the $|\alpha_{Vn}|$ and the $|\alpha_{Vp}|$.

In future, the Electron-Ion Collider in the USA (EIC) \cite{Accardi:2012qut} and the Electron-ion collider in China (EicC) \cite{Chen:2018wyz,Chen:2020ijn,Anderle:2021wcy}
will provide a good opportunity to study the near-threshold vector meson photoproduction by exploiting the virtual photon flux.
The vector meson photoproduction experiments at EIC and EicC will be further test the VMD model
and will also strengthen our understanding on the properties of hadronic interactions.

\begin{acknowledgments}
This work is supported by the Strategic Priority Research Program of Chinese Academy of Sciences
under the Grant No. XDB34030301, the National Natural Science Foundation of China No. 12005266 and
Guangdong Major Project of Basic and Applied Basic Research No. 2020B0301030008.
\end{acknowledgments}

\bibliographystyle{apsrev4-1}
\bibliography{refs.bib}

\begin{thebibliography}{27}%
\makeatletter
\providecommand \@ifxundefined [1]{%
 \@ifx{#1\undefined}
}%
\providecommand \@ifnum [1]{%
 \ifnum #1\expandafter \@firstoftwo
 \else \expandafter \@secondoftwo
 \fi
}%
\providecommand \@ifx [1]{%
 \ifx #1\expandafter \@firstoftwo
 \else \expandafter \@secondoftwo
 \fi
}%
\providecommand \natexlab [1]{#1}%
\providecommand \enquote  [1]{``#1''}%
\providecommand \bibnamefont  [1]{#1}%
\providecommand \bibfnamefont [1]{#1}%
\providecommand \citenamefont [1]{#1}%
\providecommand \href@noop [0]{\@secondoftwo}%
\providecommand \href [0]{\begingroup \@sanitize@url \@href}%
\providecommand \@href[1]{\@@startlink{#1}\@@href}%
\providecommand \@@href[1]{\endgroup#1\@@endlink}%
\providecommand \@sanitize@url [0]{\catcode `\\12\catcode `\$12\catcode
  `\&12\catcode `\#12\catcode `\^12\catcode `\_12\catcode `\%12\relax}%
\providecommand \@@startlink[1]{}%
\providecommand \@@endlink[0]{}%
\providecommand \url  [0]{\begingroup\@sanitize@url \@url }%
\providecommand \@url [1]{\endgroup\@href {#1}{\urlprefix }}%
\providecommand \urlprefix  [0]{URL }%
\providecommand \Eprint [0]{\href }%
\providecommand \doibase [0]{http://dx.doi.org/}%
\providecommand \selectlanguage [0]{\@gobble}%
\providecommand \bibinfo  [0]{\@secondoftwo}%
\providecommand \bibfield  [0]{\@secondoftwo}%
\providecommand \translation [1]{[#1]}%
\providecommand \BibitemOpen [0]{}%
\providecommand \bibitemStop [0]{}%
\providecommand \bibitemNoStop [0]{.\EOS\space}%
\providecommand \EOS [0]{\spacefactor3000\relax}%
\providecommand \BibitemShut  [1]{\csname bibitem#1\endcsname}%
\let\auto@bib@innerbib\@empty
\bibitem [{\citenamefont {Gell-Mann}\ and\ \citenamefont
  {Zachariasen}(1961)}]{Gell-Mann:1961jim}%
  \BibitemOpen
  \bibfield  {author} {\bibinfo {author} {\bibfnamefont {M.}~\bibnamefont
  {Gell-Mann}}\ and\ \bibinfo {author} {\bibfnamefont {F.}~\bibnamefont
  {Zachariasen}},\ }\href {\doibase 10.1103/PhysRev.124.953} {\bibfield
  {journal} {\bibinfo  {journal} {Phys. Rev.}\ }\textbf {\bibinfo {volume}
  {124}},\ \bibinfo {pages} {953} (\bibinfo {year} {1961})}\BibitemShut
  {NoStop}%
\bibitem [{\citenamefont {Strakovsky}\ \emph {et~al.}(2015)\citenamefont
  {Strakovsky} \emph {et~al.}}]{Strakovsky:2014wja}%
  \BibitemOpen
  \bibfield  {author} {\bibinfo {author} {\bibfnamefont {I.~I.}\ \bibnamefont
  {Strakovsky}} \emph {et~al.},\ }\href {\doibase 10.1103/PhysRevC.91.045207}
  {\bibfield  {journal} {\bibinfo  {journal} {Phys. Rev. C}\ }\textbf {\bibinfo
  {volume} {91}},\ \bibinfo {pages} {045207} (\bibinfo {year} {2015})},\
  \Eprint {http://arxiv.org/abs/1407.3465} {arXiv:1407.3465 [nucl-ex]}
  \BibitemShut {NoStop}%
\bibitem [{\citenamefont {Strakovsky}\ \emph
  {et~al.}(2020{\natexlab{a}})\citenamefont {Strakovsky}, \citenamefont
  {Epifanov},\ and\ \citenamefont {Pentchev}}]{Strakovsky:2019bev}%
  \BibitemOpen
  \bibfield  {author} {\bibinfo {author} {\bibfnamefont {I.}~\bibnamefont
  {Strakovsky}}, \bibinfo {author} {\bibfnamefont {D.}~\bibnamefont
  {Epifanov}}, \ and\ \bibinfo {author} {\bibfnamefont {L.}~\bibnamefont
  {Pentchev}},\ }\href {\doibase 10.1103/PhysRevC.101.042201} {\bibfield
  {journal} {\bibinfo  {journal} {Phys. Rev. C}\ }\textbf {\bibinfo {volume}
  {101}},\ \bibinfo {pages} {042201} (\bibinfo {year} {2020}{\natexlab{a}})},\
  \Eprint {http://arxiv.org/abs/1911.12686} {arXiv:1911.12686 [hep-ph]}
  \BibitemShut {NoStop}%
\bibitem [{\citenamefont {Strakovsky}\ \emph
  {et~al.}(2020{\natexlab{b}})\citenamefont {Strakovsky}, \citenamefont
  {Pentchev},\ and\ \citenamefont {Titov}}]{Strakovsky:2020uqs}%
  \BibitemOpen
  \bibfield  {author} {\bibinfo {author} {\bibfnamefont {I.~I.}\ \bibnamefont
  {Strakovsky}}, \bibinfo {author} {\bibfnamefont {L.}~\bibnamefont
  {Pentchev}}, \ and\ \bibinfo {author} {\bibfnamefont {A.}~\bibnamefont
  {Titov}},\ }\href {\doibase 10.1103/PhysRevC.101.045201} {\bibfield
  {journal} {\bibinfo  {journal} {Phys. Rev. C}\ }\textbf {\bibinfo {volume}
  {101}},\ \bibinfo {pages} {045201} (\bibinfo {year} {2020}{\natexlab{b}})},\
  \Eprint {http://arxiv.org/abs/2001.08851} {arXiv:2001.08851 [hep-ph]}
  \BibitemShut {NoStop}%
\bibitem [{\citenamefont {Pentchev}\ and\ \citenamefont
  {Strakovsky}(2021)}]{Pentchev:2020kao}%
  \BibitemOpen
  \bibfield  {author} {\bibinfo {author} {\bibfnamefont {L.}~\bibnamefont
  {Pentchev}}\ and\ \bibinfo {author} {\bibfnamefont {I.~I.}\ \bibnamefont
  {Strakovsky}},\ }\href {\doibase 10.1140/epja/s10050-021-00364-4} {\bibfield
  {journal} {\bibinfo  {journal} {Eur. Phys. J. A}\ }\textbf {\bibinfo {volume}
  {57}},\ \bibinfo {pages} {56} (\bibinfo {year} {2021})},\ \Eprint
  {http://arxiv.org/abs/2009.04502} {arXiv:2009.04502 [hep-ph]} \BibitemShut
  {NoStop}%
\bibitem [{\citenamefont {Wang}\ \emph
  {et~al.}(2022{\natexlab{a}})\citenamefont {Wang}, \citenamefont {Zeng},\ and\
  \citenamefont {Strakovsky}}]{Wang:2022xpw}%
  \BibitemOpen
  \bibfield  {author} {\bibinfo {author} {\bibfnamefont {X.-Y.}\ \bibnamefont
  {Wang}}, \bibinfo {author} {\bibfnamefont {F.}~\bibnamefont {Zeng}}, \ and\
  \bibinfo {author} {\bibfnamefont {I.~I.}\ \bibnamefont {Strakovsky}},\ }\href
  {\doibase 10.1103/PhysRevC.106.015202} {\bibfield  {journal} {\bibinfo
  {journal} {Phys. Rev. C}\ }\textbf {\bibinfo {volume} {106}},\ \bibinfo
  {pages} {015202} (\bibinfo {year} {2022}{\natexlab{a}})},\ \Eprint
  {http://arxiv.org/abs/2205.07661} {arXiv:2205.07661 [hep-ph]} \BibitemShut
  {NoStop}%
\bibitem [{\citenamefont {Wang}\ \emph
  {et~al.}(2022{\natexlab{b}})\citenamefont {Wang}, \citenamefont {Zeng},
  \citenamefont {Wang},\ and\ \citenamefont {Zhang}}]{Wang:2022zwz}%
  \BibitemOpen
  \bibfield  {author} {\bibinfo {author} {\bibfnamefont {X.-Y.}\ \bibnamefont
  {Wang}}, \bibinfo {author} {\bibfnamefont {F.}~\bibnamefont {Zeng}}, \bibinfo
  {author} {\bibfnamefont {Q.}~\bibnamefont {Wang}}, \ and\ \bibinfo {author}
  {\bibfnamefont {L.}~\bibnamefont {Zhang}},\ }\href@noop {} {\  (\bibinfo
  {year} {2022}{\natexlab{b}})},\ \Eprint {http://arxiv.org/abs/2206.09170}
  {arXiv:2206.09170 [nucl-th]} \BibitemShut {NoStop}%
\bibitem [{\citenamefont {Titov}\ \emph {et~al.}(2007)\citenamefont {Titov},
  \citenamefont {Nakano}, \citenamefont {Date},\ and\ \citenamefont
  {Ohashi}}]{Titov:2007xb}%
  \BibitemOpen
  \bibfield  {author} {\bibinfo {author} {\bibfnamefont {A.~I.}\ \bibnamefont
  {Titov}}, \bibinfo {author} {\bibfnamefont {T.}~\bibnamefont {Nakano}},
  \bibinfo {author} {\bibfnamefont {S.}~\bibnamefont {Date}}, \ and\ \bibinfo
  {author} {\bibfnamefont {Y.}~\bibnamefont {Ohashi}},\ }\href {\doibase
  10.1103/PhysRevC.76.048202} {\bibfield  {journal} {\bibinfo  {journal} {Phys.
  Rev. C}\ }\textbf {\bibinfo {volume} {76}},\ \bibinfo {pages} {048202}
  (\bibinfo {year} {2007})},\ \Eprint {http://arxiv.org/abs/hep-ph/0703227}
  {arXiv:hep-ph/0703227} \BibitemShut {NoStop}%
\bibitem [{\citenamefont {Workman}\ and\ \citenamefont
  {Others}(2022)}]{Workman:2022ynf}%
  \BibitemOpen
  \bibfield  {author} {\bibinfo {author} {\bibfnamefont {R.~L.}\ \bibnamefont
  {Workman}}\ and\ \bibinfo {author} {\bibnamefont {Others}} (\bibinfo
  {collaboration} {Particle Data Group}),\ }\href {\doibase
  10.1093/ptep/ptac097} {\bibfield  {journal} {\bibinfo  {journal} {PTEP}\
  }\textbf {\bibinfo {volume} {2022}},\ \bibinfo {pages} {083C01} (\bibinfo
  {year} {2022})}\BibitemShut {NoStop}%
\bibitem [{\citenamefont {Ishikawa}\ \emph {et~al.}(2020)\citenamefont
  {Ishikawa} \emph {et~al.}}]{Ishikawa:2019rvz}%
  \BibitemOpen
  \bibfield  {author} {\bibinfo {author} {\bibfnamefont {T.}~\bibnamefont
  {Ishikawa}} \emph {et~al.},\ }\href {\doibase 10.1103/PhysRevC.101.052201}
  {\bibfield  {journal} {\bibinfo  {journal} {Phys. Rev. C}\ }\textbf {\bibinfo
  {volume} {101}},\ \bibinfo {pages} {052201} (\bibinfo {year} {2020})},\
  \Eprint {http://arxiv.org/abs/1904.02797} {arXiv:1904.02797 [nucl-ex]}
  \BibitemShut {NoStop}%
\bibitem [{\citenamefont {Dietz}\ \emph {et~al.}(2015)\citenamefont {Dietz}
  \emph {et~al.}}]{CBELSATAPS:2015wwn}%
  \BibitemOpen
  \bibfield  {author} {\bibinfo {author} {\bibfnamefont {F.}~\bibnamefont
  {Dietz}} \emph {et~al.} (\bibinfo {collaboration} {CBELSA/TAPS}),\ }\href
  {\doibase 10.1140/epja/i2015-15006-3} {\bibfield  {journal} {\bibinfo
  {journal} {Eur. Phys. J. A}\ }\textbf {\bibinfo {volume} {51}},\ \bibinfo
  {pages} {6} (\bibinfo {year} {2015})}\BibitemShut {NoStop}%
\bibitem [{\citenamefont {Chang}\ \emph {et~al.}(2010)\citenamefont {Chang}
  \emph {et~al.}}]{LEPS:2009nuw}%
  \BibitemOpen
  \bibfield  {author} {\bibinfo {author} {\bibfnamefont {W.~C.}\ \bibnamefont
  {Chang}} \emph {et~al.} (\bibinfo {collaboration} {LEPS}),\ }\href {\doibase
  10.1016/j.physletb.2009.12.051} {\bibfield  {journal} {\bibinfo  {journal}
  {Phys. Lett. B}\ }\textbf {\bibinfo {volume} {684}},\ \bibinfo {pages} {6}
  (\bibinfo {year} {2010})},\ \Eprint {http://arxiv.org/abs/0907.1705}
  {arXiv:0907.1705 [nucl-ex]} \BibitemShut {NoStop}%
\bibitem [{\citenamefont {Zyla}\ \emph {et~al.}(2020)\citenamefont {Zyla} \emph
  {et~al.}}]{ParticleDataGroup:2020ssz}%
  \BibitemOpen
  \bibfield  {author} {\bibinfo {author} {\bibfnamefont {P.~A.}\ \bibnamefont
  {Zyla}} \emph {et~al.} (\bibinfo {collaboration} {Particle Data Group}),\
  }\href {\doibase 10.1093/ptep/ptaa104} {\bibfield  {journal} {\bibinfo
  {journal} {PTEP}\ }\textbf {\bibinfo {volume} {2020}},\ \bibinfo {pages}
  {083C01} (\bibinfo {year} {2020})}\BibitemShut {NoStop}%
\bibitem [{\citenamefont {Mibe}\ \emph {et~al.}(2005)\citenamefont {Mibe} \emph
  {et~al.}}]{LEPS:2005hax}%
  \BibitemOpen
  \bibfield  {author} {\bibinfo {author} {\bibfnamefont {T.}~\bibnamefont
  {Mibe}} \emph {et~al.} (\bibinfo {collaboration} {LEPS}),\ }\href {\doibase
  10.1103/PhysRevLett.95.182001} {\bibfield  {journal} {\bibinfo  {journal}
  {Phys. Rev. Lett.}\ }\textbf {\bibinfo {volume} {95}},\ \bibinfo {pages}
  {182001} (\bibinfo {year} {2005})},\ \Eprint
  {http://arxiv.org/abs/nucl-ex/0506015} {arXiv:nucl-ex/0506015} \BibitemShut
  {NoStop}%
\bibitem [{\citenamefont {Chang}\ \emph {et~al.}(2008)\citenamefont {Chang}
  \emph {et~al.}}]{Chang:2007fc}%
  \BibitemOpen
  \bibfield  {author} {\bibinfo {author} {\bibfnamefont {W.~C.}\ \bibnamefont
  {Chang}} \emph {et~al.},\ }\href {\doibase 10.1016/j.physletb.2007.11.009}
  {\bibfield  {journal} {\bibinfo  {journal} {Phys. Lett. B}\ }\textbf
  {\bibinfo {volume} {658}},\ \bibinfo {pages} {209} (\bibinfo {year}
  {2008})},\ \Eprint {http://arxiv.org/abs/nucl-ex/0703034}
  {arXiv:nucl-ex/0703034} \BibitemShut {NoStop}%
\bibitem [{\citenamefont {Koike}\ and\ \citenamefont
  {Hayashigaki}(1997)}]{Koike:1996ga}%
  \BibitemOpen
  \bibfield  {author} {\bibinfo {author} {\bibfnamefont {Y.}~\bibnamefont
  {Koike}}\ and\ \bibinfo {author} {\bibfnamefont {A.}~\bibnamefont
  {Hayashigaki}},\ }\href {\doibase 10.1143/PTP.98.631} {\bibfield  {journal}
  {\bibinfo  {journal} {Prog. Theor. Phys.}\ }\textbf {\bibinfo {volume}
  {98}},\ \bibinfo {pages} {631} (\bibinfo {year} {1997})},\ \Eprint
  {http://arxiv.org/abs/nucl-th/9609001} {arXiv:nucl-th/9609001} \BibitemShut
  {NoStop}%
\bibitem [{\citenamefont {Gao}\ \emph {et~al.}(2001)\citenamefont {Gao},
  \citenamefont {Lee},\ and\ \citenamefont {Marinov}}]{Gao:2000az}%
  \BibitemOpen
  \bibfield  {author} {\bibinfo {author} {\bibfnamefont {H.}~\bibnamefont
  {Gao}}, \bibinfo {author} {\bibfnamefont {T.~S.~H.}\ \bibnamefont {Lee}}, \
  and\ \bibinfo {author} {\bibfnamefont {V.}~\bibnamefont {Marinov}},\ }\href
  {\doibase 10.1103/PhysRevC.63.022201} {\bibfield  {journal} {\bibinfo
  {journal} {Phys. Rev. C}\ }\textbf {\bibinfo {volume} {63}},\ \bibinfo
  {pages} {022201} (\bibinfo {year} {2001})},\ \Eprint
  {http://arxiv.org/abs/nucl-th/0010042} {arXiv:nucl-th/0010042} \BibitemShut
  {NoStop}%
\bibitem [{\citenamefont {Strakovsky}\ \emph {et~al.}(2021)\citenamefont
  {Strakovsky}, \citenamefont {Briscoe}, \citenamefont {Pentchev},\ and\
  \citenamefont {Schmidt}}]{Strakovsky:2021vyk}%
  \BibitemOpen
  \bibfield  {author} {\bibinfo {author} {\bibfnamefont {I.~I.}\ \bibnamefont
  {Strakovsky}}, \bibinfo {author} {\bibfnamefont {W.~J.}\ \bibnamefont
  {Briscoe}}, \bibinfo {author} {\bibfnamefont {L.}~\bibnamefont {Pentchev}}, \
  and\ \bibinfo {author} {\bibfnamefont {A.}~\bibnamefont {Schmidt}},\ }\href
  {\doibase 10.1103/PhysRevD.104.074028} {\bibfield  {journal} {\bibinfo
  {journal} {Phys. Rev. D}\ }\textbf {\bibinfo {volume} {104}},\ \bibinfo
  {pages} {074028} (\bibinfo {year} {2021})},\ \Eprint
  {http://arxiv.org/abs/2108.02871} {arXiv:2108.02871 [hep-ph]} \BibitemShut
  {NoStop}%
\bibitem [{\citenamefont {Baillie}\ \emph {et~al.}(2012)\citenamefont {Baillie}
  \emph {et~al.}}]{CLAS:2011qvj}%
  \BibitemOpen
  \bibfield  {author} {\bibinfo {author} {\bibfnamefont {N.}~\bibnamefont
  {Baillie}} \emph {et~al.} (\bibinfo {collaboration} {CLAS}),\ }\href
  {\doibase 10.1103/PhysRevLett.108.142001} {\bibfield  {journal} {\bibinfo
  {journal} {Phys. Rev. Lett.}\ }\textbf {\bibinfo {volume} {108}},\ \bibinfo
  {pages} {142001} (\bibinfo {year} {2012})},\ \bibinfo {note} {[Erratum:
  Phys.Rev.Lett. 108, 199902 (2012)]},\ \Eprint
  {http://arxiv.org/abs/1110.2770} {arXiv:1110.2770 [nucl-ex]} \BibitemShut
  {NoStop}%
\bibitem [{\citenamefont {Tkachenko}\ \emph {et~al.}(2014)\citenamefont
  {Tkachenko} \emph {et~al.}}]{CLAS:2014jvt}%
  \BibitemOpen
  \bibfield  {author} {\bibinfo {author} {\bibfnamefont {S.}~\bibnamefont
  {Tkachenko}} \emph {et~al.} (\bibinfo {collaboration} {CLAS}),\ }\href
  {\doibase 10.1103/PhysRevC.89.045206} {\bibfield  {journal} {\bibinfo
  {journal} {Phys. Rev. C}\ }\textbf {\bibinfo {volume} {89}},\ \bibinfo
  {pages} {045206} (\bibinfo {year} {2014})},\ \bibinfo {note} {[Addendum:
  Phys.Rev.C 90, 059901 (2014)]},\ \Eprint {http://arxiv.org/abs/1402.2477}
  {arXiv:1402.2477 [nucl-ex]} \BibitemShut {NoStop}%
\bibitem [{\citenamefont {Armstrong}\ \emph
  {et~al.}(2017{\natexlab{a}})\citenamefont {Armstrong} \emph
  {et~al.}}]{Armstrong:2017zcm}%
  \BibitemOpen
  \bibfield  {author} {\bibinfo {author} {\bibfnamefont {W.~R.}\ \bibnamefont
  {Armstrong}} \emph {et~al.},\ }\href@noop {} {\  (\bibinfo {year}
  {2017}{\natexlab{a}})},\ \Eprint {http://arxiv.org/abs/1708.00835}
  {arXiv:1708.00835 [nucl-ex]} \BibitemShut {NoStop}%
\bibitem [{\citenamefont {Armstrong}\ \emph
  {et~al.}(2017{\natexlab{b}})\citenamefont {Armstrong} \emph
  {et~al.}}]{Armstrong:2017wfw}%
  \BibitemOpen
  \bibfield  {author} {\bibinfo {author} {\bibfnamefont {W.}~\bibnamefont
  {Armstrong}} \emph {et~al.},\ }\href@noop {} {\  (\bibinfo {year}
  {2017}{\natexlab{b}})},\ \Eprint {http://arxiv.org/abs/1708.00888}
  {arXiv:1708.00888 [nucl-ex]} \BibitemShut {NoStop}%
\bibitem [{\citenamefont {Armstrong}\ \emph
  {et~al.}(2017{\natexlab{c}})\citenamefont {Armstrong} \emph
  {et~al.}}]{Armstrong:2017zqr}%
  \BibitemOpen
  \bibfield  {author} {\bibinfo {author} {\bibfnamefont {W.}~\bibnamefont
  {Armstrong}} \emph {et~al.},\ }\href@noop {} {\  (\bibinfo {year}
  {2017}{\natexlab{c}})},\ \Eprint {http://arxiv.org/abs/1708.00891}
  {arXiv:1708.00891 [nucl-ex]} \BibitemShut {NoStop}%
\bibitem [{\citenamefont {Accardi}\ \emph {et~al.}(2016)\citenamefont {Accardi}
  \emph {et~al.}}]{Accardi:2012qut}%
  \BibitemOpen
  \bibfield  {author} {\bibinfo {author} {\bibfnamefont {A.}~\bibnamefont
  {Accardi}} \emph {et~al.},\ }\href {\doibase 10.1140/epja/i2016-16268-9}
  {\bibfield  {journal} {\bibinfo  {journal} {Eur. Phys. J. A}\ }\textbf
  {\bibinfo {volume} {52}},\ \bibinfo {pages} {268} (\bibinfo {year} {2016})},\
  \Eprint {http://arxiv.org/abs/1212.1701} {arXiv:1212.1701 [nucl-ex]}
  \BibitemShut {NoStop}%
\bibitem [{\citenamefont {Chen}(2018)}]{Chen:2018wyz}%
  \BibitemOpen
  \bibfield  {author} {\bibinfo {author} {\bibfnamefont {X.}~\bibnamefont
  {Chen}},\ }\bibfield  {booktitle} {\emph {\bibinfo {booktitle} {{Proceedings,
  26th International Workshop on Deep Inelastic Scattering and Related Subjects
  (DIS 2018): Port Island, Kobe, Japan, April 16-20, 2018}}},\ }\href {\doibase
  10.22323/1.316.0170} {\bibfield  {journal} {\bibinfo  {journal} {PoS}\
  }\textbf {\bibinfo {volume} {DIS2018}},\ \bibinfo {pages} {170} (\bibinfo
  {year} {2018})},\ \Eprint {http://arxiv.org/abs/1809.00448} {arXiv:1809.00448
  [nucl-ex]} \BibitemShut {NoStop}%
\bibitem [{\citenamefont {Chen}\ \emph {et~al.}(2020)\citenamefont {Chen},
  \citenamefont {Guo}, \citenamefont {Roberts},\ and\ \citenamefont
  {Wang}}]{Chen:2020ijn}%
  \BibitemOpen
  \bibfield  {author} {\bibinfo {author} {\bibfnamefont {X.}~\bibnamefont
  {Chen}}, \bibinfo {author} {\bibfnamefont {F.-K.}\ \bibnamefont {Guo}},
  \bibinfo {author} {\bibfnamefont {C.~D.}\ \bibnamefont {Roberts}}, \ and\
  \bibinfo {author} {\bibfnamefont {R.}~\bibnamefont {Wang}},\ }\href {\doibase
  10.1007/s00601-020-01574-0} {\bibfield  {journal} {\bibinfo  {journal} {Few
  Body Syst.}\ }\textbf {\bibinfo {volume} {61}},\ \bibinfo {pages} {43}
  (\bibinfo {year} {2020})},\ \Eprint {http://arxiv.org/abs/2008.00102}
  {arXiv:2008.00102 [hep-ph]} \BibitemShut {NoStop}%
\bibitem [{\citenamefont {Anderle}\ \emph {et~al.}(2021)\citenamefont {Anderle}
  \emph {et~al.}}]{Anderle:2021wcy}%
  \BibitemOpen
  \bibfield  {author} {\bibinfo {author} {\bibfnamefont {D.~P.}\ \bibnamefont
  {Anderle}} \emph {et~al.},\ }\href@noop {} {\  (\bibinfo {year} {2021})},\
  \Eprint {http://arxiv.org/abs/2102.09222} {arXiv:2102.09222 [nucl-ex]}
  \BibitemShut {NoStop}%
\end{thebibliography}%

\end{document}